\documentclass[amsmath,amssymb,a4paper,twocolumn,prb]{revtex4}
\usepackage{txfonts}

\usepackage{graphicx}

\makeatletter
\def\bib@device#1#2{}
\makeatother

\begin{document}
\makeatletter 　　
\newcommand{\rmnum}[1]{\romannumeral #1} 　　
\newcommand{\Rmnum}[1]{\expandafter\@slowromancap\romannumeral #1@}
\makeatother
\title{Intrinsic anisotropy of thermal conductance in graphene nanoribbons}
\author{Yong Xu, Xiaobin Chen, Bing-Lin Gu, and Wenhui Duan\footnote{Author to whom correspondence should be
addressed. E-mail address: dwh@phys.tsinghua.edu.cn}}
\affiliation{Center for Advanced Study and Department of Physics,
Tsinghua University, Beijing 100084, China}


\begin{abstract}
Thermal conductance of graphene nanoribbons (GNRs) with the width
varying from 0.5 to 35 nm is systematically investigated using
nonequilibrium Green's function method. Anisotropic thermal
conductance is observed with the room temperature thermal
conductance of zigzag GNRs up to $\sim$30\% larger than that of
armchair GNRs. At room temperature, the anisotropy is found to
disappear until the width is larger than 100 nm. This intrinsic
anisotropy originate from different boundary condition at ribbon
edges, and can be used to tune thermal conductance, which have
important implications for the applications of GNRs in
nanoelectronics and thermoelectricity.
\end{abstract}

\maketitle

Graphene nanoribbon (GNR) is believed to be a promising candidate
for future nanoelectronic and spintronic
devices.~\cite{geim,duan,duan2,louie-nature} In comparison with
intensive research on the electronic transport, the study of the
thermal transport in GNRs is not satisfactory until now, despite its
importance for nanoelectronics. At present the chip-level power
density in integrated circuit is on the order of 100 W/cm$^2$,
similar to that of nuclear reactor. Increasing circuit density
further would induce the exponential increase of power density, so
thermal management on individual nanoscale devices becomes vital to
ensure stable operation of integrated circuit.~\cite{pop} This
indicates a full understanding of thermal transport property of GNRs
is critical for developing any practical graphene-based devices.

It was previously demonstrated that basic device building blocks
can be constructed from GNRs with different widths and edge
shapes, which dominates their electronic properties.~\cite{duan}
For the evaluation of the overall performance of GNR devices, it
is highly important to explore thermal transport characteristics
of GNRs with different edge shapes and widths. In fact, a recent
work found anisotropic thermal conductance in silicon nanowires
(SiNWs): $\langle 110 \rangle$ SiNW exhibits a room temperature
thermal conductance 50\% and 75\% larger than $\langle 100
\rangle$ and $\langle 111 \rangle$ SiNWs.~\cite{jauho} In
contrast, systematical investigation on whether GNRs with
different edge shapes give anisotropic thermal conductance is
still lacking, despite recent progress in the calculations of
thermal conductances of GNRs.~\cite{yamamoto-prb,lan} A classical
molecular dynamics simulation studied thermal conductance for the
1.5 nm wide GNRs with different edges,~\cite{hu} but the thermal
conductance obtained obviously exceeds the upper ballistic
bounds.~\cite{hu_note,mingo-prl}

In this work, we systematically investigate thermal conductance of
GNRs with different widths and edge shapes, and find that thermal
conductance of GNRs is primarily proportional to the ribbon width
except for GNRs narrower than 2 nm, and more interestingly, exhibits
strong anisotropy. Typically, thermal conductance of zigzag GNRs
(ZGNRs) are 20\% $\sim$ 30\% larger than that of armchair GNRs
(AGNRs) at room temperature as the ribbon width ranges from 0.5 to 2
nm. Such intrinsic anisotropy decreases with increasing width, but
is still larger than 10\% even when the width is as large as 35 nm,
and thus may have significant effect on the performance of the GNR
devices.

\begin{figure}
\centering\includegraphics[width=0.45\textwidth]{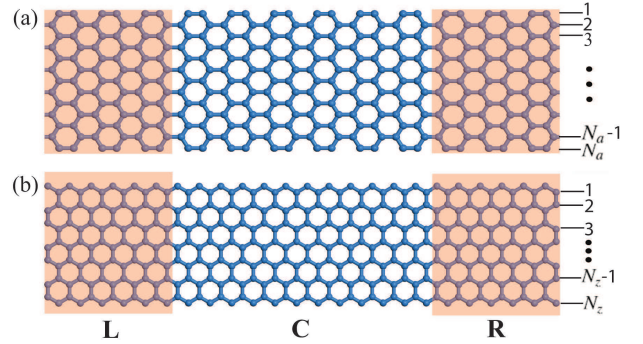}
\caption{Schematic illustration of two-probe transport system,
where the center transport region (C) is connected to left (L) and
right (R) semi-infinite thermal leads. (a) $N_a$-AGNR, (b)
$N_z$-ZGNR. }\label{fig1}
\end{figure}

Nonequilibrium Green's function (NEGF) method is used to study
thermal conductance of GNRs. This method is powerful to treat many
body problems when interactions are weak, and can exactly deal
with ballistic thermal transport, which gives the maximum thermal
conductance of a
material.~\cite{yamamoto-prl,mingo-prb,xu,jswang-epj} Since the
phonon mean free path is very long in graphene ($\sim$ 775 nm at
room temperature)~\cite{ghosh}, the thermal transport is nearly
ballistic in pristine GNRs, and thus ballistic thermal conductance
is studied in this work. First of all, the second-generation
reactive empirical bond order (REBOII) potential,~\cite{brenner}
which was proved to give phonon modes of GNRs compatible with
density functional theory calculations,~\cite{andescuren} is
employed to calculate force constants. Then, Green's function and
the transmission function $\Xi (\omega)$ of phonon can be
calculated for the two-probe transport system (shown in
Fig.~\ref{fig1}), and finally thermal conductance $\sigma$ is
obtained by Landauer formula.~\cite{jswang-epj} Since GNRs with
larger width ($W$) will have more phonon transport channels and
thus larger thermal conductance, the scaled thermal conductance,
defined as thermal conductance per unit area ($\sigma / S$), is
introduced to describe thermal transport properties for materials
with different widths. Herein, the cross sectional area $S$ is
defined to be $S = W\delta$, where $\delta = 0.335$ nm is chosen
as the layer separation in graphite. Following the conventional
notation, $N_a$-AGNR ($N_z$-ZGNR) denotes an AGNR (a ZGNR) with
$N_a$ ($N_z$) carbon dimer lines (zigzag carbon chains) across the
ribbon width, as shown in Fig.~\ref{fig1}. For the sake of
comparison, armchair carbon nanotubes (ACNTs) and zigzag carbon
nanotubes (ZCNTs) are also considered and their widths refer to
their circumferences.

\begin{figure}
\centering\includegraphics[width=0.45\textwidth]{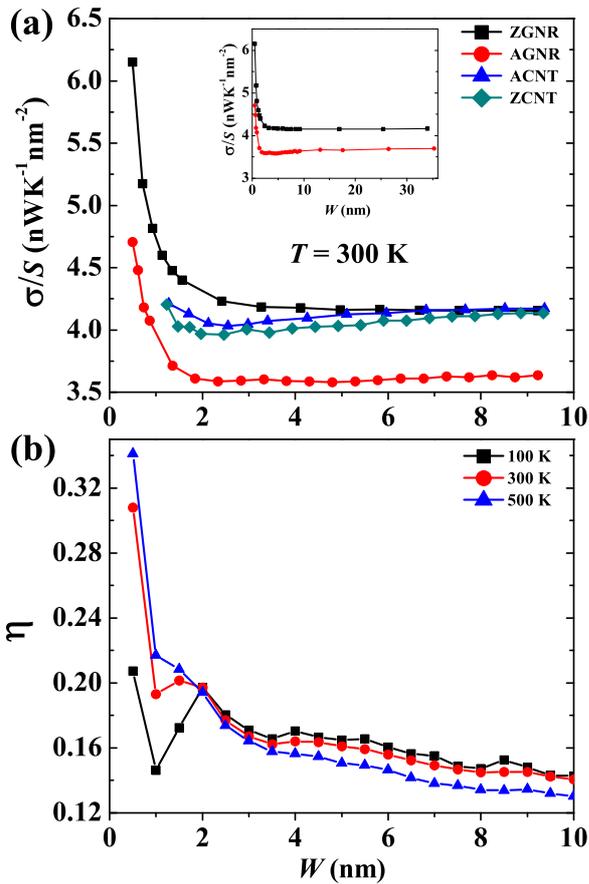}
\caption{(Color online) (a) The scaled thermal conductance
($\sigma / S$) at 300 K versus width ($W$) for ZGNRs (black
square), AGNRs (red circle), ACNTs (blue triangle) and ZCNTs (cyan
diamond). The inset shows $\sigma / S$ for ZGNRs and AGNRs with
the width varying from 0.5 to 35 nm. (b) The anisotropy factor
($\eta$) versus width ($W$) at 100 K (black square), 300 K (red
circle) and 500 K (blue triangle). The lines are drawn to guide
the eyes.}\label{fig2}
\end{figure}

Fig.~\ref{fig2}(a) shows the scaled thermal conductance ($\sigma /
S$) at 300 K versus width for GNRs.  It is interesting to see that
GNRs exhibit different properties with respect to carbon nanotubes
(CNTs). ACNTs have nearly the same scaled thermal conductance as
ZCNTs, consistent with previous theoretical
results\cite{mingo-prl}. Moreover, the scaled thermal conductance
of CNTs changes only slightly as the width changes. In one word,
the scaled thermal conductance of CNTs is almost independent of
their chirality and size. This, however, is not the case in GNRs,
where significant size effect and anisotropy of thermal
conductance are observed. For GNRs narrower than 2 nm, the scaled
thermal conductance decreases rapidly as the width increases;
while for wider GNRs, it changes very slowly. More important,
ZGNRs have significantly larger scaled thermal conductance than
AGNRs, indicating a strong anisotropic thermal transport in GNRs.
For example, when the width is larger than 2 nm, the scaled
thermal conductance of ZGNRs at 300 K is around 4.2 nW/K/nm$^2$,
which is close to that of CNTs, while that of AGNRs is about 3.6
nW/K/nm$^2$. The room temperature thermal conductance per unit
area of wide GNRs is of the same order as that of pristine SiNWs
($\sim$ 1 nW/K/nm$^2$)~\cite{jauho}. However, graphene has much
larger room temperature thermal conductivity ($\sim 5 \times 10^4$
W/K/m)~\cite{balandin} than SiNWs ($\sim$ 6 W/K/m for a 22 nm
diameter SiNW)~\cite{li}, because the phonon mean free path is
much longer in graphene than in SiNWs.

To give a quantitative description of the anisotropic thermal
conductance, we further define an anisotropy factor for the
thermal conductance as $\eta = [(\sigma / S)_{\rm ZGNR}/(\sigma /
S)_{\rm AGNR}] - 1$. Fig.~\ref{fig2}(b) shows the anisotropy
factor $\eta$ of GNRs with the width ranging from 0.5 to 10 nm at
different temperatures (100, 300 and 500 K). It can be seen that
narrower GNRs generally exhibit stronger anisotropy. However, the
variation of the anisotropy factor is irregular at all
temperatures when $W$ is less than 2 nm. At 300 K, the anisotropy
factor changes irregularly from 31\% to 20\% when the width varies
from 0.5 nm to 2 nm, and decrease monotonously to 14\% as the
width increases to 10 nm. When the temperature changes from 100 K
to 500 K, the anisotropy factor changes little except for very
narrow GNRs. Moreover, the anisotropy factor will decrease rapidly
when the temperature decreases from 100 K (the data is not shown
here).

One can expect the anisotropy factor of GNRs will decrease to zero
when the width is large enough, since thermal conductance is
isotropic in graphene sheets.~\cite{saito} It is important to
determine the critical width at which the anisotropy of thermal
conductance disappears. We have extended our calculations to the
ribbon width up to 35 nm. As shown in the inset of
Fig.~\ref{fig2}(a), the scaled thermal conductance varies very
slowly with the width for wide GNRs. The anisotropy factor of 35
nm wide GNR still has a value of 13\% at 300 K. Obviously, a
direct calculation of thermal conductance at critical width using
the NEGF method is beyond our computational capability. So we
apply linear regression to fit the data from 4 to 35 nm for ZGNRs
and AGNRs, and find that the anisotropy may disappear when $W\sim$
140 nm at 300 K.

What is the origin of anisotropic thermal conductance in GNRs? It
was once attributed to different phonon scattering rates at the
edges.~\cite{hu} For ideal structures without defects, however,
there will be no boundary scattering at all, since the boundary
structure is incorporated into the phonon modes. In fact, our
results show the anisotropy still exists even without edge
scattering. It was previously revealed that the anisotropic thermal
conductance of SiNWs can be reproduced from the anisotropic phonon
structure of bulk silicon~\cite{jauho}. Differently, graphene, the
bulk counterpart of GNRs, is isotropic in thermal conductance. This
suggests that for GNRs, boundary condition at edges instead of the
bulk property is the origin of anisotropy.

\begin{figure}
\centering\includegraphics[width=0.45\textwidth]{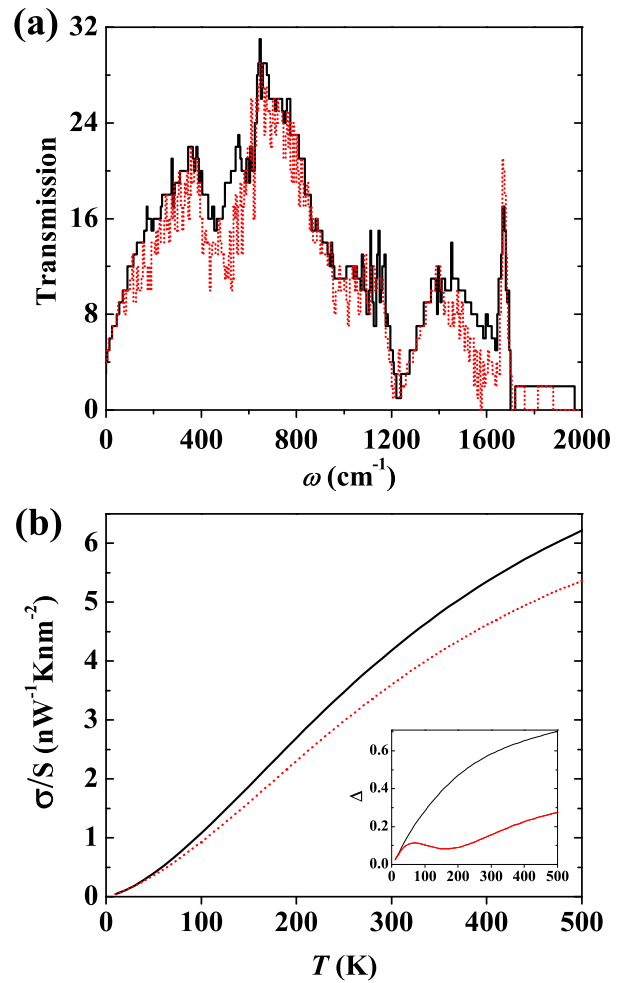}
\caption{(Color online) (a) Transmission function versus phonon
frequency ($\omega$) and (b) the scaled thermal conductance
($\sigma / S$) versus temperature ($T$) for 16-ZGNR (black solid
line) and 28-AGNR (red dotted line). The inset shows the decrease
of scaled thermal conductance ($\Delta$) induced by fixing edge
atoms. }\label{fig3}
\end{figure}

The effect of the boundary condition on thermal conductance is
rather complicated. Since lattice vibration is a collective
behavior, almost every phonon modes will change if the boundary
condition varies. Moreover, thermal conductance is always
contributed by many phonon modes, and thus the total effect can not
be demonstrated by focusing on particular phonon modes, but should
be analyzed from the whole transmission. Fig.~\ref{fig3}(a) shows
the phonon transmission function for 16-ZGNR and 28-AGNR, which have
a similar width ($W \sim 3.3$ nm). Only negligible difference in
transmission is observed at low frequency region ($\omega < 100$
cm$^{-1}$), implying that the anisotropy disappears when the
temperature is lower than 25 K. This indicates that the boundary
condition has little influence on low frequency phonon modes. In
contrast, obvious effect of boundary condition can be seen at high
frequency region (especially at 400 $\sim$ 600 cm$^{-1}$ and 1400
$\sim$ 1650 cm$^{-1}$): the transmission of 28-AGNR is always lower
than that of 16-ZGNR. In principle, transmission function of
ballistic transport is determined by the number and the dispersion
of phonon bands. More dispersive phonon bands would give larger
transmission. Since the two GNRs have nearly the same number of
atoms per unit volume, the discrepancy should not be caused by
different number of phonon bands. Comparing with zigzag edge,
armchair edge indeed gives more localized lattice vibrations and
less dispersive phonon bands, resulting in the anisotropic thermal
conductance in GNRs.

The dependence of the scaled thermal conductance on temperature is
illustrated in Fig.~\ref{fig3}(b) for 16-ZGNR and 28-AGNR. As
expected, the conductance increases monotonously with increasing
temperature. Evidently, 16-ZGNR always has larger scaled thermal
conductance than 28-AGNR. The anisotropy factor is small at low
temperatures and keeps around 16\% from 100 K to 500 K (data is not
shown). Since the scaled thermal conductance only slightly varies
with the width when the width is larger than 2 nm, it can be used to
determine the maximum phonon thermal conductance at different
temperatures for wide GNRs. Thermal conductance of a ZGNR (an AGNR)
($W > $ 2 nm) can be estimated through multiplying the scaled
thermal conductance of 16-ZGNR (28-AGNR) by the cross sectional
area.

To further understand the effect of boundary condition on
anisotropic thermal conductance, we artificially impose fixed
boundary condition on the above two GNRs (i.e., 16-ZGNR and
28-AGNR) by setting the mass of edge atoms to $1.0 \times 10^7$
atomic mass unit in the calculation. Fixing edge atoms causes a
decrease of scaled thermal conductance for both GNRs, as shown in
Fig.~\ref{fig3}(b). The scaled thermal conductance of 16-ZGNR
decreases noticeably, especially at high temperatures, whereas the
decrease of 28-AGNR is relatively small. After imposing fixed
boundary condition the anisotropy becomes smaller but does not
disappear. This shows the anisotropy is robust even in this
extreme condition, which is substantially different from the free
or periodic boundary condition generally used.

It should be noted that both phonon and electron transport can
contribute to thermal conductance in practice. The electronic
thermal conductance of gated GNRs was well discussed in
Ref.~\onlinecite{watanabe}. Since recent experiments showed that
all sub-10-nm GNRs are semiconducting,~\cite{dai} thermal
conductance in GNRs narrower than 10 nm should be mostly dominated
by phonons at all temperatures. While for GNRs of several tens of
nm wide, the contribution from electrons might become important at
low temperatures.

In conclusion, we have systematically studied thermal conductance
for ZGNRs and AGNRs with the ribbon width ranging from 0.5 to 35
nm. Similar to electronic property, thermal conductance can also
be tuned by the ribbon width and edge shape. The scaled thermal
conductance ($\sigma/S$) decreases as the width increases and
becomes nearly width independent when width is larger than 2 nm.
More interestingly, ZGNRs have 13\% $\sim$ 31\% larger thermal
conductance than AGNRs at 300 K, showing that thermal conductance
is anisotropic in GNRs. Unlike in SiNWs, the anisotropy in GNRs is
caused by different boundary condition at edge but not the bulk
phonon structure. Comparing with zigzag edge, armchair edge gives
flatter phonon bands and behaves more like a fixed boundary. The
intrinsic anisotropy found here is not limited to GNRs, but is
also expected to appear in other materials with hexagonal
structure, and is strong enough to be observable experimentally.
Our work provides useful insight into the thermal management in
GNR-based nanoelectronic and thermoelectric devices.

This work was supported by the Ministry of Science and Technology
of China (Grant No. 2006CB605105 and 2006CB0L0601), the National
Natural Science Foundation of China.



\begin{references}
\bibitem{geim} A. H. Castro Neto, F. Guinea, N. M. R. Peres, K. S.
Novoselov, and A. K. Geim, Rev. Mod. Phys. \textbf{81}, 109
(2009); and references therein.

\bibitem{louie-nature} Y. W. Son, M. L. Cohen, and S. G. Louie, Nature \textbf{444}, 347
(2006).

\bibitem{duan} Q. Yan, B. Huang, J. Yu, F. Zheng, J. Zang, J. Wu, B.-L. Gu, F. Liu, and W. Duan, Nano Lett. \textbf{7}, 1469 (2007).

\bibitem{duan2} Z. Li, H. Qian, J. Wu, B.-L. Gu, and W. Duan, Phys. Rev. Lett. \textbf{100}, 206802 (2008).

\bibitem{pop} E. Pop, S. Sinha, and K. E. Goodson, Proc. IEEE  \textbf{94}, 1587 (2006).

\bibitem{jauho} T. Markussen, A.-P. Jauho, and M. Brandbyge, Nano Lett. \textbf{8}, 3771 (2008).

\bibitem{yamamoto-prb} T. Yamamoto and K. Watanabe, Phys. Rev. B \textbf{70}, 245402
(2004).

\bibitem{lan} J. Lan, J.-S. Wang, C. K. Gan, and S. K. Chin, Phys. Rev. B
\textbf{79}, 115401 (2009).

\bibitem{hu} J. Hu, X. Ruan, and Y. P. Chen, Nano Lett. \textbf{9}, 2730 (2009).

\bibitem{hu_note} The thermal conductivity given by Ref. \onlinecite{hu} is
about 2000 W/m/K at 400 K for a 5.7 nm long, 1.5 nm wide ZGNR, and
the related thermal conductance is about 176 nW/K, much larger
than the maximum thermal conductance (about 3 nW/K) from our
ballistic phonon transport calculations. It was previously pointed
out that molecular dynamics simulation is inadequate to describe
quantum ballistic transport and can lead to the substantial
violation of the ballistic upper bounds to thermal conductance
(see Ref.~\onlinecite{mingo-prl}).

\bibitem{mingo-prl} N. Mingo and D. A. Broido, Phys. Rev. Lett. \textbf{95}, 096105 (2005).

\bibitem{yamamoto-prl} T. Yamamoto and K. Watanabe, Phys. Rev. Lett.
\textbf{96}, 255503 (2006).

\bibitem{mingo-prb} N. Mingo, Phys. Rev. B \textbf{74}, 125402 (2006).

\bibitem{xu} Y. Xu, J.-S. Wang, W. Duan, B.-L. Gu, and B. Li, Phys. Rev. B
\textbf{78}, 224303 (2008).

\bibitem{jswang-epj} J.-S. Wang, J. Wang, and J. T. L\"u, Eur. Phys. J. B \textbf{62}, 381 (2008).

\bibitem{ghosh} S. Ghosh, I. Calizo, D. Teweldebrhan, E. P. Pokatilov, D. L. Nika,
A. A. Balandin, W. Bao, F. Miao, and C. N. Lau, Appl. Phys. Lett.
\textbf{92}, 151911 (2008).

\bibitem{brenner} D. W. Brenner, O. A. Shenderova, J. A. Harrison, S. J Stuart, B. Ni,
and S. B Sinnott, J. Phys.: Condens. Matter \textbf{14}, 783 (2002).

\bibitem{andescuren} M. Vandescuren, P. Hermet, V. Meunier, L. Henrard, and Ph.
Lambin, Phys. Rev. B \textbf{78}, 195401 (2008).

\bibitem{balandin} A. A. Balandin, S. Ghosh, W. Bao, I. Calizo, D. Teweldebrhan, F. Miao,
and C. N. Lau, Nano Lett. \textbf{8}, 902 (2008).

\bibitem{li}D. Li, Y. Wu, P. Kim, L. Shi, P. Yang, and A. Majumdar, Appl.
Phys. Lett. \textbf{83}, 2934 (2003).

\bibitem{saito} K. Saito, J. Nakamura, and A. Natori, Phys. Rev. B \textbf{76}, 115409 (2007).

\bibitem{watanabe} E. Watanabe, S. Yamaguchi, J. Nakamura, and A.
Natori, Phys. Rev. B \textbf{80}, 085404 (2009).

\bibitem{dai} X. Li, X. Wang, L. Zhang, S. Lee, and H. Dai, Science \textbf{319},
1229 (2008).

\end{references}
\end{document}